# Inference of Causal Information Flow in Collective Animal Behavior

Warren M. Lord, Jie Sun, Nicholas T. Ouellette, and Erik M. Bollt

*Abstract*—Understanding and even defining what constitutes animal interactions remains a challenging problem. Correlational tools may be inappropriate for detecting communication between a set of many agents exhibiting nonlinear behavior. A different approach is to define coordinated motions in terms of an information theoretic channel of direct causal information flow. In this work, we consider time series data obtained by an experimental protocol of optical tracking of the insect species *Chironomus riparius*. The data constitute reconstructed 3-D spatial trajectories of the insects' flight trajectories and kinematics. We present an application of the optimal causation entropy (oCSE) principle to identify direct causal relationships or information channels among the insects. The collection of channels inferred by oCSE describes a network of information flow within the swarm. We find that information channels with a long spatial range are more common than expected under the assumption that causal information flows should be spatially localized. The tools developed herein are general and applicable to the inference and study of intercommunication networks in a wide variety of natural settings.

## I. Introduction

COLLECTIVELY interacting groups of social animals such as herds, schools, flocks, or crowds go by many names depending on the specific animal species. In all cases, they tend to display seemingly purposeful, coordinated group-level dynamics despite the apparent absence of leaders or directors. These coordinated group behaviors appear to emerge only from interactions between individuals, analogous to the manner in which macroscopic observables are determined by microscopic interactions in statistical physics. Thus, collective behavior has captivated a broad spectrum of researchers from many different disciplines [1]–[19].

Manuscript received May 31, 2016; revised September 30, 2016; accepted November 14, 2016. This work was supported in part by the Army Research Office under Grant W911NF-12-1-0276 (WML, JS, EMB), Grant W911NF-16-1-0081 (WML, JS, EMB), and Grant W911NF-16-1-0185 (NTO), and in part by the Simons Foundation under Grant 318812 (JS). The associate editor coordinating the review of this paper and approving it for publication was S. M. Moser.
W. M. Lord, J. Sun, and E. M. Bollt are with the Department of Mathematics, Clarkson University, Potsdam, NY 13699 USA (e-mail: lordwm@clarkson.edu; sunj@clarkson.edu; bolltem@clarkson.edu).
N. T. Ouellette is with the Department of Civil and Environmental Engineering, Stanford University, Stanford, CA 94305 USA (e-mail: nto@stanford.edu).
Digital Object Identifier 10.1109/TMBMC.2016.2632099

Making the analogy with statistical physics more concrete, it is reasonable to suggest that a deep understanding of collective group motion may arise from three parallel pursuits. We can perform a macroscopic analysis, focusing on the observed group-level behavior such as the group morphology [20] or the material-like properties [18], [21], [22]; we can perform a microscopic analysis, determining the nature of the interactions between individuals [10], [19], [23], [24]; and we can study how the microscopic interactions scale up to give rise to the macroscopic properties [25].

The third of these goals—how the microscopic individual-to-individual interactions determine the macroscopic group behavior—has arguably received the most scientific attention to date, due to the availability of simple models of collective behavior that are easy to simulate on computers, such as the classic Vicsek *et al.* [25], Reynolds [26], and Couzin *et al.* [27] models. From these kinds of studies, a significant amount is known about the nature of the emergence of macroscopic patterns and ordering in active, collective systems [28]. But in arguing that such simple models accurately describe real animal behavior, one must implicitly make the assumption that the interactions between individuals are correctly represented. Any model of interactions has two key and distinct components: a specification of the mathematical form of the interaction, and, more fundamentally, a choice as to *which* individuals interact. Given that it is difficult to extract the appropriate social interaction network from empirical measurements, models typically replace this hard-to-measure social network with the simple-to-define proximity network [29]. Thus, it is assumed that individuals interact only with other animals that are spatially nearby. No matter what species is involved, the answer to the question of whether interactions are generally limited to or dominated by spatial local neighbors has strong implications. Recently, for example, scientists studying networks have shown that introducing even a small number of long range interactions into a lattice can impart qualitative changes to the observed macroscopic behavior [30], [31]. Consequently, the question of whether flocks or swarms or herds also contain long range interactions between individuals may have important implications for the understanding of collective motion.

Efforts to move past the simple framework of assuming that the local spatial neighborhood of an individual dominates its behavior have been largely theoretical [32], [33], as it is challenging to extract the underlying interaction network from measured data. Empirical methods have often relied upon various types of correlational[1] (or other pairwise) time-series analysis [19], which by design only capture linear dependence and fail to detect the nonlinear relationships that are typical in real-world applications. An alternative paradigm would be to use information theoretic methods that are capable of detecting nonlinear dependence. Examples of such include directed information [34], [35] and the more recently proposed Transfer Entropy (TE) [36], the latter being a type of conditional mutual information designed specifically for the detection of (possibly asymmetric) information flow between two coupled units. For instance: transfer entropy analysis was used in attempt to infer leader-follower relationships among flying bats from video data [37], and a variant of it called local transfer entropy was adopted to understand information transfer among spatially nearby soldier crabs [38]. Prior to these works, local transfer entropy was also used in synthetic data of swarms to study information cascades in these systems [39]. However, such TE-based analysis also rely on *pairwise* computations, and thus cannot differentiate between direct and indirect interactions. As we have recently shown, any pairwise method, no matter how high its fidelity, will tend to overestimate the number of links in the interaction network, typically resulting in a significant number of false positives that cannot be resolved even with unlimited data [40], [41].

In this paper, we introduce a new mathematical framework based on optimal causation entropy (oCSE)[2] to reveal the detailed network of interactions between individuals in a collective animal group. In brief, Causation Entropy (CSE) is capable of detecting and describing the interactions between three and higher numbers of components [40], [41]. That is, we can describe the influence of individual $X$ on individual $Y$ conditioned on the existence of the influence $Z$ on $Y$. Thus, CSE allows us to draw conclusions quite different from what we could using pairwise interactions alone, in that information that "flows" from $X$ to $Y$ only through $Z$ is an indirect influence that would at best be misclassified as a direct influence by a pairwise measure. Built upon the concept of CSE, oCSE is an efficient, constructive algorithm to infer the network of *direct* interactions based on CSE estimates [41]. Note that a similar algorithm that uses transfer entropy was previously presented in [46]. In the application of inferring interactions among animals, oCSE requires knowledge only of the positions (or velocities or accelerations) of individuals in a group and is thus directly computable from empirical data. Because we define interactions via the information theoretic notion of the direct exchange of information as detected by uncertainty reduction, we need not make any assumptions about the spatial proximity of interacting individuals or the precise mathematical form of interaction. To demonstrate the unique utility of this oCSE network inference algorithm, we apply it to experimental measurements of the motion of individuals in mating swarms of the non-biting midge *Chironomus riparius*. In addition to showing the computability of the CSE in this data set, the oCSE approach clearly reveals that spatial proximity and interaction are not synonymous, suggesting that a deep understanding of collective behavior requires more subtle analysis of interactions than simple position-based proximity metrics.

We note that a recent paper [47] shows that under model-based assumptions regarding the information flow in a network of fish, line-of-sight type of interactions provides a better fit of the empirical fish movement data than several standard interaction protocols, including spatial nearest-neighbors. In this paper, the use of oCSE bypasses the necessity of forming model-specific assumptions including detailed physical mechanisms behind information sharing. In addition, insects provide an important case example where communications are believed to occur not only through vision, but also through the air medium (sound) [18], [48] (we mention this later in Section III) and potentially in other ways. The oCSE approach can be applied to animal species beyond fish and insects, as it virtually accounts for any type of information sharing while appropriately conditioning out indirect links.

## II. THE oCSE ALGORITHM AND STATISTICAL INFERENCE

### A. Causation Entropy and the oCSE Algorithm

We say roughly that information flows from agent $X$ to agent $Y$ if the future states of $Y$ and the past states of $X$ share information that cannot be accounted for by any other variables in the system. To make this statement precise we need to define the information associated with a variable, what it means to share information, and how to condition on other potential sources of information.

The information content descriptive of a random variable can be quantified by the Shannon entropy. Thus, if $p(x)$ is the probability that a measurement of a variable $X$ take the particular value $x$, then the uncertainty associated with that variable is defined as its Shannon entropy [49],

$$H(X) \equiv -\sum_x p(x) \log p(x), \quad (1)$$

where the summation is taken over all possible values of $X$ with the convention $p(x) \log p(x) = 0$ if $p(x) = 0$. The base of the logarithm is not important for inferring relationships between variables as long as the same base is adopted. If $X$ is a real-valued continuous random variable, we may interpret $p(x)$ as the probability density function of $X$ and use differential entropy, defined as $h(X) \equiv -\int_{-\infty}^{\infty} p(x) \log p(x)\, dx$. In this section we use the notation $H$ to mean (1); but $h(X)$ may also apply, depending on the application. Note that in the application to midge swarming, we assume that insect acceleration is a continuous variable.

When two random variables are available the information in $X$ can be subdivided into information belonging only to $X$ and information belonging to both $X$ and the other variable, $Y$. The mutual information describes the shared information between $X$ and $Y$ and is defined as [49], [50]

$$I(X; Y) = H(X) + H(Y) - H(X, Y). \quad (2)$$

---

[1] Unless otherwise noted, throughout the paper we use "correlation" to mean the commonly adopted "Pearson linear correlation" in data analysis.

[2] In this paper the terms "causal" and "causation" are used in the sense of information (as opposed to physical) causality, following the line of thought by Clive Granger and his eventual Nobel-Prize winning concept [42], [43]. Causal in this case is interpreted as "appear to have caused" (making it possible to infer from post-hoc analysis of passively observed data [41]). That is we have used these phrases to be descriptive of predicatively related in keeping with other highly popular concepts such as Granger causality and also Transfer Entropy for example. On the other hand, another line of thought leads to defining causal as "must have caused" which requires active interventions [44], [45] that are often impractical for the data analysis of large complex systems.

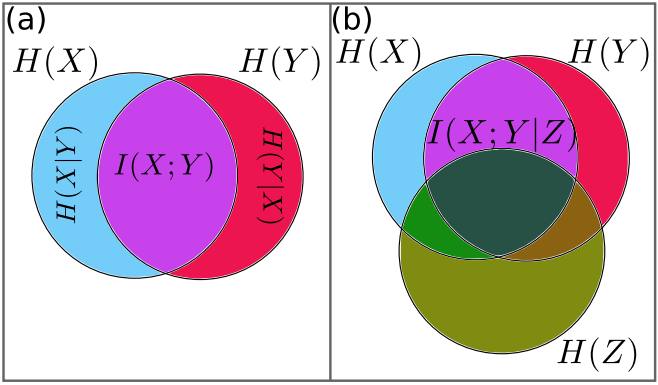
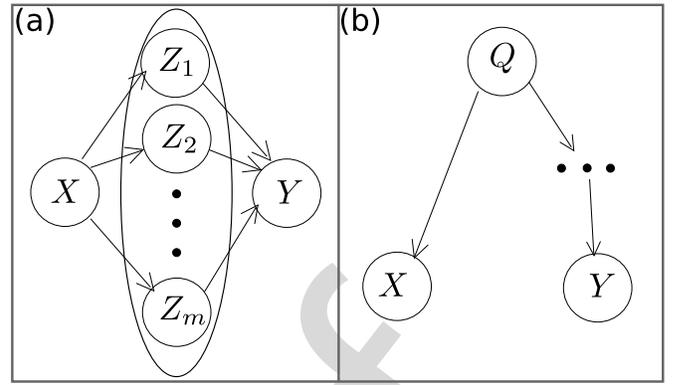

Fig. 1. Entropy and mutual information as visualized by Venn diagrams. Note that unlike a Venn diagram for sets, the Venn diagram for entropies should be interpreted with the caution that certain "areas" could be negative, and therefore conditioning has the seemingly paradoxical potential to increase mutual information (see [41] for explicit examples).

Fig. 2. Two ways that $I(X^{(t)}, Y^{(t+\tau)}|Y^{(t)})$ could be positive without an edge between $X$ and $Y$ in a discrete time dynamical process on a directed graph. (a) The variables $Z_i$ (or even a subgraph of $\{Z_i\}_{i=1}^{m}$) serve as an intermediary between $X$ and $Y$. (b) The states of $X$ and $Y$ are strongly influenced by a third source, $Q$. The "..." indicate that there might be other nodes on the path from $Q$ to $Y$ to induce the time lag $\tau$.

Here $H(X, Y)$ denotes the entropy of the joint random variable $(X, Y)$ whose measurements consist of ordered pairs, $(x, y)$. Fig. (1a) shows a Venn diagram visualization of the relation between various entropies and mutual information.

Conditioning is similar to removing part of a circle in Fig. (1a) from the picture, leaving a crescent of the other circle remaining. For example, the conditional entropy of $Y$ given $X$ [49], [50],

$$H(Y|X) = H(X, Y) - H(X), \quad (3)$$

tells how much uncertainty is associated with $Y$ given knowledge about $X$. The importance of conditional entropy for understanding swarm behavior is in finding the mutual information between two variables that is not present in a third variable. If $Z$ is another discrete variable then the conditional mutual information of $X$ and $Y$ given $Z$ is defined by

$$I(X; Y|Z) = H(X|Z) - H(X|Y, Z), \quad (4)$$

which is visualized in the information Venn diagram, Fig. (1b).

Transfer Entropy is a type of conditional mutual information [36]. If information in $X$ "flows" to $Y$ then 1) there would have to be information contained in $Y$ at a future time, say $t + \tau$, that is not explained by the state of $Y$ at time $t$, and 2) this information would be shared by $Y$ at time $t + \tau$ and $X$ at time $t$. There are three variables present, which we can label $X^{(t)}$, $Y^{(t)}$, and $Y^{(t+\tau)}$. For information to "flow" from $X$ to $Y$ over time $\tau$, then, it is necessary that $I(X^{(t)}; Y^{(t+\tau)}|Y^{(t)}) > 0$. The quantity on the left is proposed as Transfer Entropy $T_{X \to Y}$ because it reveals how much information has been transferred by the flow. This notion of an information flow, however, does not reflect how the information got there. When there are more than two insects present, information could flow from $X$ to $Y$ by going through a set of intermediaries [Fig. (2a)], or the information could have flowed to both from a third source, $Q$, but taken slightly longer to reach $Y$ than $X$ [Fig. (2b)].

Causation Entropy (CSE) is a quantity designed to detect directed information flows. If $X$, $Y$ and $Z$ are variables related to three insects then the Causation Entropy of $X$ to $Y$ given $Z$ is

$$C_{X \to Y|Z} = I\left(X^{(t)}; Y^{(t+\tau)}|Z^{(t)}\right). \quad (5)$$

In other words, $C_{X \to Y|Z}$ is the information shared between $X^{(t)}$ and $Y^{(t+\tau)}$ that is not already contained in $Z^{(t)}$. CSE as defined in (5) generalizes Transfer Entropy since the conditioning set does not necessarily contain information about the past of $Y$ [41]. With such a definition of CSE in terms of conditioning sets, the question becomes how to appropriately select the conditioning sets to reveal *direct* information flow in a network. To this end, below we review oCSE as an algorithmic approach to efficiently learn the underlying interaction network structure. A similar algorithm that uses TE can be found in [46].

The oCSE algorithm starts with an initial conditioning set and only adds as many variables as necessary [41]. This is called the aggregative or discovery phase, which is followed by a removal phase in which redundancies are removed from the set [41]. To be more specific, let $\mathcal{X} = \{X_1, X_2, \ldots, X_m\}$. Initially, let $\mathcal{Z}$ be an empty set (although in some applications prior knowledge enables one to select a non-empty initial set). On each round the variable $X_i$ is added to $\mathcal{Z}$ if

$$C_{X_i \to Y|\mathcal{Z}} = \max_{X_j \notin \mathcal{Z}} C_{X_j \to Y|\mathcal{Z}} > 0. \quad (6)$$

The discovery phase terminates when no such variable can be found from the remaining set of variables. The resulting set $\mathcal{Z}$ is possibly a superset of the variables that communicate directly with $Y$ because the value of $C_{X_i \to Y|\mathcal{Z}}$ can in fact be positive due to indirect information flow from $X_i$ to $Y$, unless $\mathcal{Z}$ contains all the other true causal components [41]. Thus, the removal phase eliminates elements from $\mathcal{Z}$ if they are redundant given other elements in $\mathcal{Z}$. On each iteration a new member of $\mathcal{Z}$, $Z_i$, is chosen and removed if and only if

$$C_{Z_i \to Y | \mathcal{Z} \setminus \{Z_i\}} = 0. \tag{7}$$

After all variables of $\mathcal{Z}$ are considered, those that remain in $\mathcal{Z}$ are the direct causal parents of $Y$, meaning that information flows directly from the elements of $\mathcal{Z}$ to $Y$. This assertion was proved by the optimal Causation Entropy Principle (oCSE) [41] that provides multiple characterizations of the set of causal parents [41]. The causal parent relationship is written $X_i \to Y$. Note that the collection of all relationships $X_i \to X_j$ forms a directed graph in which the variables are the nodes and edges represent the direction of information flow. This graph can have cycles. In fact it is quite possible that $X_i \to X_j$ and $X_j \to X_i$. In a swarm this situation would be analogous to a "dance" in which two midges were interacting and *mutually* adjusting their movements in accordance with the other insect's movement.

### B. Practical Considerations: oCSE From Time Series Data

In practice, the CSE values need to be *estimated* from data. Since CSE is expressed as a conditional mutual information, what is needed is essentially a "good", consistent estimator for conditional mutual information. Development of such estimators is an important computational and statistical problem that is of general relevance to a significant body of work in the literature. In this work we adopted a nonparametric estimator of conditional mutual information derived in [51] and [52] as an extension of the Kraskov-Strögbauer-Grassberger (KSG) [53] estimator for mutual information. Details of this estimator are provided in the Appendix.

In addition to the estimation of CSE from data, a key step in the algorithmic inference (such as oCSE) of direct causal links from data is determining whether or not the estimated CSE value $\widehat{C_{X \to Y | Z}}$ should be regarded as being strictly positive. To address this, we consider a shuffle test for the null hypothesis of $\widehat{C_{X \to Y | Z}} = 0$ [41] (also see [54]–[56]). Given time series samples $\{(x_t, y_t, z_t)\}$ of a stochastic process $(X_t, Y_t, Z_t)$, the idea is that for the estimated $\widehat{C_{X \to Y | Z}}$ to be regarded as significantly positive, it should typically "beat" (i.e., be greater than) $\widehat{C_{X' \to Y | Z}}$ where $X'$ corresponds to a surrogate data obtained by replacing $\{x_t\}$ with $\{x'_t\}$, the latter being a random permutation of the set $\{x_t\}$. By repeating the estimation with a large number of permutations, we consider $C_{X \to Y | Z}$ to be significant if $\widehat{C_{X \to Y | Z}}$ is greater than a fraction $(1 - \alpha)$ of the values of $\widehat{C_{X' \to Y | Z}}$ where $\alpha$ is a prescribed level of significance. Note, however, that such $\alpha$ in a single significance test of CSE should not be interpreted as the significance level of the directed links inferred by the entire oCSE algorithm because the inclusion/exclusion of a link typically requires multiple testings that are not necessarily independent. Consequently, the false positive rate for the entire algorithm is expected to be somewhat larger than $\alpha$. In [41] we numerically found that for a class of multivariate Gaussian distributions the false positive ratio is closely related to the $\alpha$ used in the individual tests. Exact correspondence between $\alpha$ and the significance level of the inferred links in general remains an open challenge.

Furthermore, even though the number of variables in this study is moderately small, the question of computational complexity is important for the oCSE algorithm to be utilized in the future for much larger number of insects and data points. For fixed sample size and choice of CSE estimator, each iteration of oCSE requires the computation of $O(n)$ number of CSE values. This, together with the total number of iterations, $K \leq n$, gives the computational complexity of oCSE to be $O(Kn)$ for the inference of causal parents of a single node/variable in the network. For sparse networks, the value of $K$ typically equals the degree of the target node [41] and thus the computational complexity to infer the entire network is expected to be $O(m)$ where $m < n^2$ is the number of links in the network. This is to be contrasted with classical combinatorial-search based algorithms such as the PC algorithm [57] for which the computational complexity of inferring the entire network is $O(n^2(n-1)^{k-1}/(k-1)!)$ where $k$ is the maximum degree. Due to the incremental, non-combinatorial nature, the number of data samples required by oCSE to achieve a given level of inference accuracy (for a given directed pair) does not depend much on the network size $n$, but rather, the density of links [41].

## III. APPLICATION OF THE oCSE ALGORITHM TO SWARMING INSECTS

Both to demonstrate the types of information that can be gleaned from the direct interaction networks inferred by the oCSE approach and to show that such networks are computable for real empirical data sets that contain noise and other non-idealities, we apply oCSE to empirical measurements of swarming insects. Here, we briefly describe the experimental methodology, including the insect husbandry procedures and data acquisition system, and then show the results of the oCSE computation. These results enable us to compare and contrast spatially nearest neighbors with direct causal neighbors.

### A. Experimental Methods

Many different species of insects in the order Diptera exhibit swarming as a part of their mating ritual [58], and such swarms are a well studied, canonical example of collective behavior. Swarms are also an excellent model system for testing the oCSE algorithm: since swarms are internally disordered and show little overall pattern or correlation [59], it is difficult to tell by eye which individuals, if any, are interacting.

Here, we apply the oCSE algorithm to data collected from the observation of swarms of the non-biting midge *Chironomus riparius* under controlled laboratory conditions. Details of our insect husbandry procedures and experimental protocols have been reported in detail elsewhere [60], [61], so we described them only briefly here. Our breeding colony of midges is kept in a cubic enclosure measuring 91 cm on a side; temperature and humidity are controlled via laboratory climate-control systems. Midge larvae develop in 9 open tanks, each containing 7 L of oxygenated, dechlorinated water and a cellulose substrate into which the larvae can burrow.

Adult midges live in the same enclosure, typically sitting on the floor or walls when they are not swarming. The entire enclosure is illuminated from above by a light source that provides 16 hours of light in each 24-hour period. When the light turns on and off, male midges spontaneously form swarms. We encourage swarm nucleation and position the swarms in the enclosure by means of a "swarm marker" (here, a 32×32 cm piece of black cloth) placed on the floor of the enclosure. The number of midges participating in each swarming event is uncontrolled; we have observed swarms consisting of as few as one or two midges and as many as nearly 100 [62].

To quantify the kinematics of the midges' flight patterns, we reconstruct the time-resolved trajectory of each individual midge via automated optical particle tracking. The midge motion during swarming is recorded by three Point Grey Flea3 digital cameras, which capture 1 megapixel images at a rate of 100 frames per second (fast enough to resolve even the acceleration of the midges [60]). The three cameras are arranged in a horizontal plane outside the midge enclosure with angular separations of roughly 45°. Bright light can disrupt the natural swarming behavior of the midges; thus, we illuminate them in the near infrared, which the midges cannot see but that the cameras can detect. In each 2D image on each camera, midge positions are determined by simple image segmentation followed by the computation of intensity-weighted centroids. These 2D positions were then combined together into 3D world coordinates via stereomatching, using a pinhole model for each camera and calibrating via Tsai's method [63]. To match the individual 3D positions together into trajectories, we used a fully automated predictive particle-tracking method originally developed to study highly turbulent fluid flows [64]. Occasionally, tracks will be broken into partial segments, due to mistakes in stereoimaging or ambiguities in tracking; to join these segments together into long trajectories, we used Xu's method of re-tracking in a six-dimensional position-velocity space [65]. After tracks were constructed, accurate velocities and accelerations were computed by convolving the trajectories with a smoothing and differentiating kernel [62]. The final data set for each swarming event therefore consists of time series of the 3D position and its time derivatives for each midge.

### B. Inferring Insect Interactions Using oCSE: Choice of Variables, Parameters, and Conditioning

After acquiring the experimental data, several decisions need to be made with respect to the choice of variables, parameters, and conditioning in the oCSE algorithm, in order to produce meaningful results that are interpretable.

The data sets used here contained empirical measurements from 126 distinct swarming events with varying numbers of participating individuals. For each swarm, time series of position, velocity, and acceleration were collected for each individual insect. We narrowed down the data by considering the collection of long (> 1 second) disjoint time intervals in which the corresponding data contains exactly the same number (≥ 5) of insects in each such interval. In other words we restricted our studies to data sets where the same "actors" were at play throughout the window of study.

Since the datasets include the spatial trajectories, the available variables include position, velocity, acceleration (all 3D), or any reasonable function of one or more of these, for example functions that project onto individual coordinate axes. We used the 3D acceleration data as opposed to position or velocity, for two reasons. One is that accelerations are often interpreted as "social forces" in the animal motion literature [61], and so seemed the most suited form of data for investigating interactions. The other reason is that the acceleration showed less autocorrelation than either the position or velocity data.

Next, given that the motion of the insects is not stationary, we expect the causal influences to vary over time even within each experiment. For this reason, we apply oCSE to infer causal networks from data defined in relatively small time windows instead of the entire time span of each experiment. We choose the time window size to be 1 second, corresponding to 100 data samples (at a sampling frequency of $100Hz$), which seems to be the minimal window size that produce relatively continuous-in-time causal networks.

In addition, we seek to infer causal influences from other midges *beyond* the influence of a midge to itself. To ensure that the self-influence is properly accounted for, we modified the starting point of the oCSE algorithm as described in Section II to be $\mathcal{Z} = \{Y\}$ (for a given midge $Y$), and always keep $Y \in \mathcal{Z}$ in both the discovery and removal phases.

Furthermore, the time-lag $\tau$ is chosen to be $\tau = 0.05$ seconds (5 time steps) based on biological considerations. In particular, we observed that [61] the midges tend to travel in a straight line for (usually less or equal to) 0.1 seconds before making a sudden acceleration over the next few frames, as if there were gathering and processing information during their straight flight before reacting. Fourier analysis reveals that the most important frequency for this acceleration is 1 acceleration per 0.1 seconds. Thus, the time lag should be smaller than 0.1 seconds in order to capture these accelerations. However, making $\tau$ too small reduces the amount of useful information comparing against noise. The choice of $\tau = 0.05$ seconds achieves a reasonable compromise.

Finally, the estimator of CSE (see the Appendix) requires a choice of the parameter $k$. In this study we fixed $k = 4$ for all computations (too small of a $k$ gives estimates with high variance), noting that a number of papers offer heuristics for choosing $k$ [66], [67] as a function of sample size. We chose hypothesis tests with significance level $\alpha = 0.01$ in both the forward (discovery) and backward (removal) phases of the oCSE algorithm. In theory, $\alpha$ should control the sparsity of the desired graph but we were unable to confirm this numerically. We typically ran 1000 trials per hypothesis test.

### C. Results

The most basic result of applying the oCSE algorithm is the determination of the *direct* causal links between individuals in the swarm. Although the data contains many suitable time series describing the motion of more than 30 swarming midges, Figs. 3 and 4 describe a small swarm of 5 midges for the purpose of illustrating the application of oCSE to finding direct causal links. In Fig. 3, we show four consecutive snapshots of these links from the perspective of a single insect (labelled as "1") over a period of 2 seconds. In the first snapshot, panel (a), midge 1 is identified as being influenced by midges 2 and 4; that is, it is receiving information from them. Notice that in the second snapshot, (b), the link from 2 to 1 has been lost, but 1 is still receiving information from 4. Perhaps because it moved closer, 1 is also receiving information from 3 in the second snapshot. By panel (c), 1 seems to have noticed 5, but by the final snapshot this link has been lost.

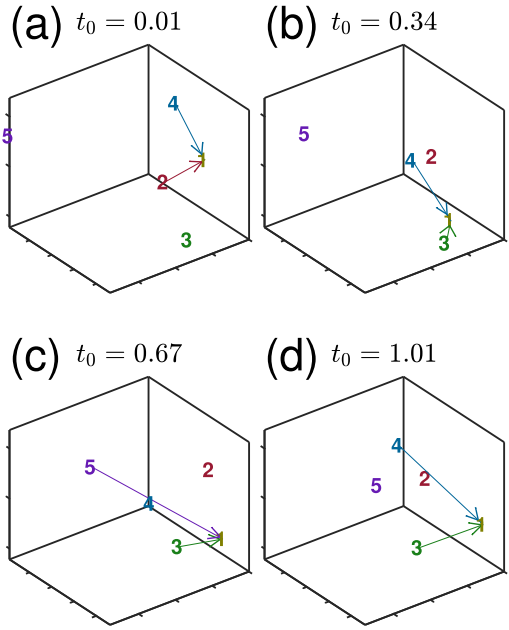

Fig. 3. Information flow from the perspective of midge 1. Only information flows into midge 1 (i.e., direct causal links to 1) are depicted, and are represented by the solid lines. Each panel corresponds to an oCSE computation using 1 second (100 frames) of data. The positions of the midges are given by their initial positions, $t_0$ during the interval. The time lag—see the definition of CSE in Eq. (5)—is $\tau = 0.05$ seconds. The initial time $t_0 = 0.01$ of panel (a) is chosen to be the point when insect 5 becomes observable.

Such transient interactions are reminiscent of those we described earlier using a different (time-series-based) measure [19]. In that case, we had hypothesized that the primary purpose of such interactions was for the registration of the gender of other midges in the swarm, since the biological purpose of swarming in this species is mating. A similar process may be at work here, and midge 1 may have, for example, successfully identified midge 2 after the first snapshot so that further information transfer was unnecessary.

In addition to studying only the information flows into a single insect, we can look at the entire directed graph returned by the oCSE algorithm. In Fig. 4, we show this full directed graph for the two initial conditions corresponding to frames (a) and (b) of Fig. 3. The direction of an edge is given by its color, where the source of the information flow determines the color.

One easy set of statistics to read off of these graphs are the in and out degrees. The in-degree of a node is the number of edges pointing to that node and the out-degree is the number of edges with one end at the node but pointing to a different node. In these plots, the out-degree of a node is the number of edges that are plotted in the same color as the node and the in-degree is the number of edges with an end at the node but which are plotted in a different color. So, for instance, in Fig. (4a), the in-degree of midge 1 is 2 (verifying the computation used to create Fig. (3a)), and the out-degree is 3.

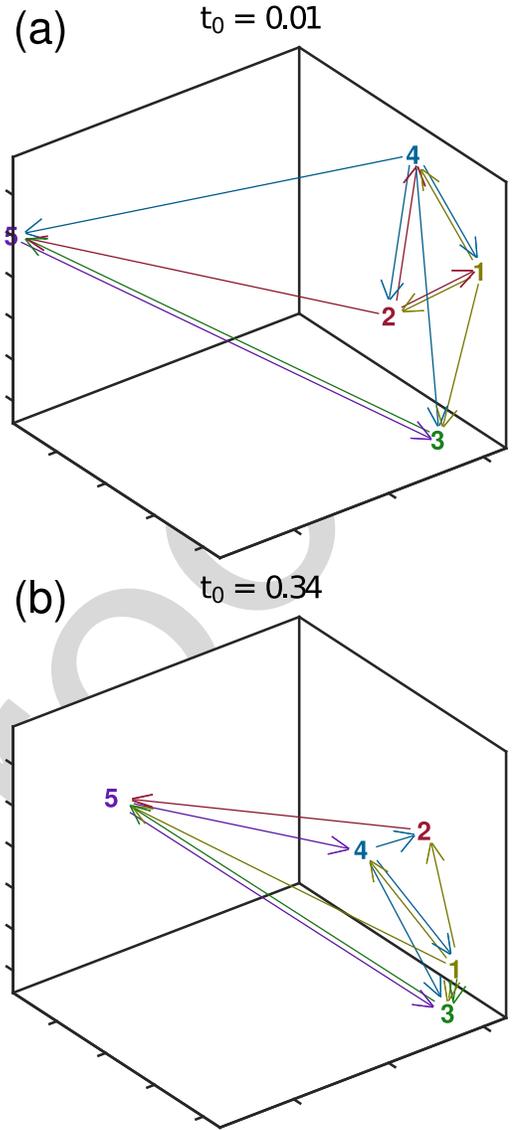

Fig. 4. Directed graphs of all inferred information flows corresponding to the first two panels of Fig. 3. Each edge represents a flow of information and takes on the color of the information source. The parameters used in the oCSE algorithm are the same as in Fig. 3: $\tau = 0.05$, each computation uses 100 frames of data, and the positions are determined by the first frame.

The average in-degree (which is always equal to the average out-degree) is $12/5 = 2.4$ in both (a) and (b) of Fig. 4. In general, we found that the average number of causal neighbors per midge to be in between 2 and 3, and such number does not seem to change as a function of the swarm/network size in our experiments. In fact, for more than 95% of the cases over all analyzed swarming data (many of which contain $> 5$ midges), the maximum number of causal neighbors is always less or equal to 5. These findings suggest that a typical midge pays attention to a relatively constant number of other midges at any particular time. This hypothesis is lent more credence by noting that the in-degree of every individual midge is the same in (a) and (b).

The out-degrees are much more variable, however. Biologically speaking, out-degrees may give information on which midges are the most important, in the sense that if a midge has a high out-degree then others seem to be reacting to the motions of this midge. In Fig. (4a), although the most spatially central node, midge 2, has an out-degree of 3, so does midge 1, which is not as spatially central. Furthermore, midge 4 has the largest out-degree with every other midge paying attention to 4. So, although midge 2 is the most spatially central node, we say that midge 4 has the highest "degree centrality". A similar analysis can be carried out on panel (b) showing that at $t_0 = 0.34$, midge 4 is now the most spatially central, but node 1 has the highest degree centrality. Some statistics that give more detailed information about centrality are eigenvector centrality and betweenness centrality.

Rather than attempting to comprehensively apply all available graph analysis methods, we give two simple observations and refer the reader to [68] for more on the analysis of graphs. In Fig. (4a) the subset of midges $\{3, 5\}$ forms a "sink" for information. Although 3 and 5 are gathering information from many midges, they are apparently unable to send that information to any other midges than 3 and 5. If one were to code edges in an "adjacency matrix" of 1's representing edges and 0's represent the lack of an edge, this feature corresponds to the adjacency matrix being reducible. The set $\{3, 5\}$ generates the only non-trivial subgraph closed under inclusion of all out-going edges. A less restrictive analysis that is similar in flavor is community detection [68], but this type of analysis is usually reserved for larger graphs.

Again in Fig. (4a), the edges linking $\{1, 2, 4\}$ generate a triangle in which information can flow in both directions around the triangle. This is a special relationship between three nodes called a 3-clique. It should be compared with $\{2, 4, 5\}$ in Fig. (4b) in which information flows only in one direction. In swarms with many other midges it is conceivable that most randomly picked triplets $\{a, b, c\}$ would have no triangle between them. The density of different types of triangles in a set is quantified by the clustering coefficient. Social networks tend to have much higher clustering as measured by clustering coefficients than technological networks and networks whose edges are determined randomly [69].

Because both Figs. 3 and 4 show that the configuration of causal links can and does change in time, it is reasonable to ask about the temporal variability and stability of the oCSE results: if the links switch seemingly at random from time step to time step, then the results would be unintelligibly unreliable. To check the stability and reliability of the oCSE results, we computed the causal links for sets of overlapping time intervals, as shown in Fig. 5 for a particular example. Although there is occasionally some drop-out of links from one instant to the next, the overall results of the algorithm are clearly stable and more-or-less continuous in time; and we conjecture that the discontinuously dropped causal links could in fact be restored, if necessary, by improving the tracking

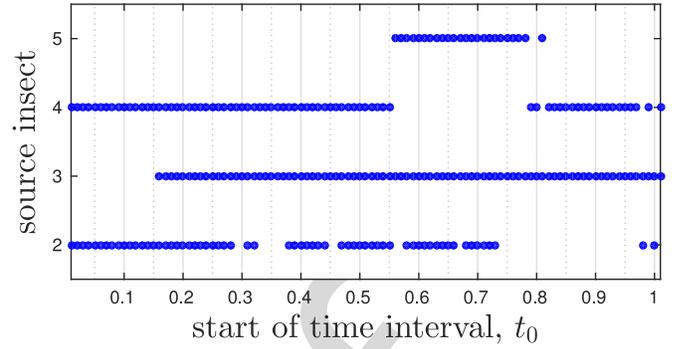

Fig. 5. Evolution of a causal neighborhood of a single midge over one second following the trajectories of 5 midges. This figure gives a more (temporally) detailed look at the information flows depicted in Fig. 3 at the expense of spatial information. Again, midge 1 is the target individual; that is, we are inferring causal links toward midge 1. Edges from $j$ to 1 are replaced by a point at $(t_0, j)$ for the appropriate $t_0$. The inputs and parameters are the same as those used to create Fig. 3. Long stretches of symbols or blank space demonstrate that the oCSE algorithm is robust to changes in the spatial configuration and kinematics of the swarm.

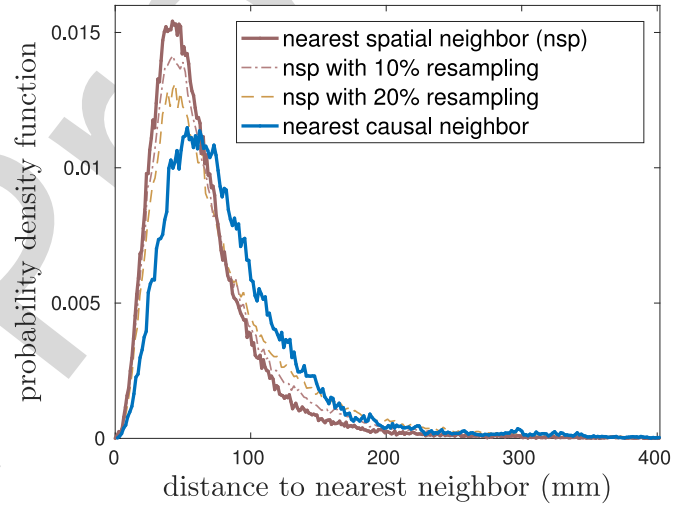

Fig. 6. Probability density functions for distance from a midge to its: nearest spatial neighbor versus nearest causal neighbor. The distances used for constructing these pdfs are obtained by selecting a single midge randomly at each time slice and computing the minimal spatial distance to: other midges (spatial nearest neighbor distance), or causal neighbors. Furthermore, to account for the potential errors from statistical estimation and testing in causal inference, we also construct and plot the "resampled" pdfs of spatial nearest distances, by replacing a fixed fraction (10% or 20%) of these distances with distances to randomly chosen other midges.

accuracy or by some post-processing step or some combination of both.

Finally, as noted above, it is intriguing to note that midges connected via causal links are not always the spatially closest to each other. Although a full characterization and complete understanding of the distinction between spatial proximity and causal information flow is beyond the scope of the present paper, we can at least describe at a statistical and macroscopic level the difference between those by measuring the probability density functions (pdfs) of the distance between nearest spatial neighbors and nearest causal neighbors. These pdfs are plotted in Fig. 6 and show that as compared to simply calculating minimum distances for given time slices from the raw data, the restriction to causal neighbors by the oCSE algorithm significantly shifts the typical distance between neighbors to larger values. Since choosing any neighbor other than the spatial nearest neighbor will always shift the distribution to the right, the nearest causal neighbor distribution should be compared to the distributions for nearest spatial neighbor in which 10% or 20% of the neighbors are chosen randomly. These distributions can be interpreted as the distribution of distances to neighbors under weakened nearest neighbor assumptions. So for instance, the distribution for distance to nearest spatial neighbor with 20% resampling is the distribution that would occur if each insect followed a rule of 4 out of 5 times following their closest neighbor and the remainder of times randomly following a random neighbor. The distribution of distances to nearest causal neighbors seems unlikely even under this loose interpretation of the nearest neighbor rule. From these observations (also see Fig. 6), it seems that a typical insect often moves independently of its spatially closest neighbors.

## IV. Conclusion

Information theoretic methods show great promise for the analysis of biological data. In particular, the oCSE algorithm allows the inference of directed graphs of information flow from time series data. The introduction of the oCSE algorithm allows a wider variety of questions about collective animal motion to be addressed. In particular, this enables us to pose the question of which individual animals are *directly* interacting with which other individual animals. Such analysis opens up possibilities and questions well beyond standard group analysis, which may typically be based on mean-field behaviors. Specifically, here we have suggested that it is possible to consider which animal may be acting as a center of attention, and how this scenario may change in a time-varying network of information flow of influence. Further we have noted that contrary to standard models of swarming and group behaviors, these information flow networks allow that the influential animals for each individual animal may not necessarily be the closest spatial neighbors. With these new analysis tools, we hope that a wide variety of new questions can be posed and studied directly from experimental data, moving beyond current phenomenological models that are validated in terms of interpretive rather than observational understanding.

Despite some preliminary and interesting findings obtained from the application of oCSE to time series data gathered from insect motion, several fundamental challenges remain. First, we considered acceleration in this study mainly to reduce the effect of autocorrelation which would otherwise dominate other available information. It will be interesting to investigate, for a different dataset or experiment, other choice of variables such as position or velocity or some combination of these. Secondly, different insect species have different memory mechanisms and response time. Our choice of the time delay ($\tau = 0.05$ seconds) might be suitable for this particular dataset, but it will be useful to consider other values and even ask whether there is an optimal choice for a given species. It is also possible to introduce heterogeneous time delays given that the processing time and response time might differ vastly for certain animal species. Furthermore, improved estimators of CSE (or conditional mutual information) would likely enhance the reliability of inference especially when there is limited amount of data comparing to the size of the sample space. It will be particularly useful to develop estimators that, together with significance testing, minimize the inference errors even if the estimator itself might not be optimal in inferring the individual CSE values. On the other hand, with the rapid development of experimental conditions and data collection methods, we expect more data to be available in the future, likely to lead to more accurate inference. As with any statistical inference, testing of significance is central but an exact test remains to be developed for causality inference. Finally, it is worth pointing out that even with abundant data and excellent estimation, the meaning and inference of causality still requires some level of assumptions and cannot be simply addressed beyond doubt.

## Appendix
### Estimation of CSE From Data

In practice, formula (5) defining CSE must be estimated from time series data. By time lagging the time series for the effect variable, the estimation of $C_{X \to Y|Z}$ can be reduced to the estimation of the conditional mutual information, or to the estimation of mutual information when the conditioning set is empty.

A straightforward way to estimate entropy and related quantities is by binning (histogram) methods, which effectively estimate a probability density $p(x)$ by counting the frequency of sampled points falling in a constant-size region around that point. Although conceptually simple to understand and easy to implement, such binning methods have been shown to suffer from slow convergence, especially for multivariate data sets, as they scale badly with the embedding dimension. Faster convergence can be achieved by nonparametric estimators based on $k$-nearest neighbor (knn) statistics. The basic idea is to estimate the density at a given point using distance to the $k$ nearest neighbors rather than neighbors falling within a constant-size neighborhood. For mutual information, we adopt the Kraskov-Strögbauer-Grassberger (KSG) estimator, which was shown to be data efficient (with $k = 1$ the estimator resolve structures down to the smallest possible scales), adaptive (the resolution is higher where data are more numerous), and to have minimal bias (the bias is mainly due to nonuniformity of the density at the smallest resolved scale, giving typical systematic errors that scale as functions of $k/N$ for $N$ points) [53]. The KSG estimator can be extended for the estimation of conditional mutual information. One such extension was recently proposed by Frenzel and Pompe [52] and independently by Vejmelka and Paluš [51]. Consider $n$ independent samples $\{w_1, w_2, \ldots, w_n\}$ of the joint random variable $W = (X, Y, Z)$ where $w_i = (x_i, y_i, z_i)$. The estimate of $I(X; Y|Z)$ is given by

$$I(X; Y|Z) = \psi(k) - \langle \psi(n_{xz} + 1) + \psi(n_{yz} + 1) - \psi(n_z + 1) \rangle. \quad \text{(A.1)}$$

Here $\langle \cdot \rangle$ denotes the average over the samples and $\psi(t) = \Gamma'(t)/\Gamma(t)$ is the digamma function. For fixed value of $k$, let $\epsilon(i)$ denotes the distance from $w_i \equiv (x_i, y_i, z_i)$ to its $k$th nearest neighbor, where distance is measured as the max-norm in the joint space, $\|w_i - w_j\|_{xyz} = \max\{\|x_i - x_j\|_x, \|y_i - y_j\|_y, \|z_i - z_j\|_z\}$, and the norms used in the subspaces can be arbitrary but often time the max norm is used (which is our choice for this paper) as well. From this we obtain

- $n_{xz}(i)$: number of points $(x_j, z_j)$ ($j \neq i$) with $\|(x_j, z_j) - (x_i, z_i)\|_{xz} = \max\{\|x_j - x_i\|_x, \|z_j - z_i\|_z\} < \epsilon(i)$
- $n_{yz}(i)$: number of points $(y_j, z_j)$ ($j \neq i$) with $\|(y_j, z_j) - (y_i, z_i)\|_{yz} = \max\{\|y_j - y_i\|_y, \|z_j - z_i\|_z\} < \epsilon(i)$
- $n_z(i)$: number of points $z_j$ ($j \neq i$) with $\|z_j - z_i\|_z < \epsilon(i)$

Several recent papers have focused on reducing the finite-sample bias of the KSG type of estimators [70]–[72] or developing other types of estimators [73].

## Acknowledgment

The authors thank Dr. Samuel Stanton from the ARO Complex Dynamics and Systems Program for his ongoing and continuous support.

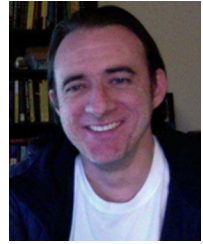

**Warren M. Lord** received the B.S. degree from Texas Christian University in 2011 and the M.S. degree in applied mathematics from the University of Colorado Boulder in 2014. He is currently pursuing the Ph.D. degree in mathematics from Clarkson University, Potsdam, NY, USA.

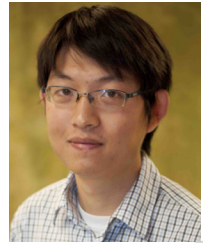

**Jie Sun** received the B.Sc. degree in physics and the B.Ec. degree in finance from Shanghai Jiao Tong University, Shanghai, China, in 2006, and the M.Sc. and Ph.D. degrees in mathematics from Clarkson University, Potsdam, NY, USA, in 2007 and 2009, respectively. He spent two and a half years as a Post-Doctoral Researcher with the Department of Physics and Astronomy, Northwestern University, Evanston, IL, USA, and the Department of Molecular Biology, Princeton University, Princeton, NJ, USA. He is currently an Assistant Professor with the Department of Mathematics, Clarkson University, with a joint courtesy appointment with the Department of Physics and the Department of Computer Science. His research covers a broad range of topics in applied mathematics and data science, including the structure and dynamics of complex networks, causality inference, computational inverse problems, information theory, and big data visualization and analytics.

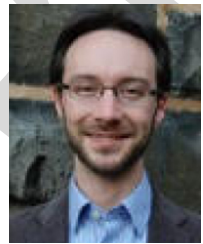

**Nicholas T. Ouellette** received the B.A. degree in physics and computer science from Swarthmore College, Swarthmore, PA, USA, in 2002, and the Ph.D. degree in physics from Cornell University, Ithaca, NY, USA, in 2006. He was a Post-Doctoral Fellow with the Max Planck Institute for Dynamics and Self-Organization, Göttingen, Germany, in 2006, and the Department of Physics, Haverford College, Haverford, PA, USA, from 2007 to 2008. He is currently an Associate Professor of Civil and Environmental Engineering, Stanford University, Stanford, CA, USA. He was with the faculty of Mechanical Engineering and Materials Science, Yale University, New Haven, CT, USA. He is broadly interested in the behavior of complex systems far from equilibrium, with ongoing research projects in fluid dynamics, granular mechanics, and collective behavior in biology.

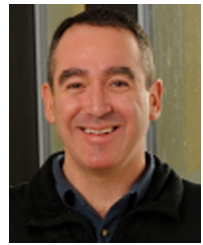

**Erik M. Bollt** received the degree in applied mathematics from the University of California at Berkeley, CA, USA, and the University of Colorado at Boulder, CO, USA. He is currently the W. Jon Harrington Professor of Mathematics with the Department of Mathematics, Clarkson University, Potsdam, NY, USA. He was with the faculty of Mathematics with the U.S. Naval Academy, Annapolis, MD, USA, and with the U.S. Military Academy, West Point, NY, USA. He studies the intersection of applied dynamical systems and data analytics. Broadly this includes complex and networked systems as informed by data and experimental or remote observations by data methods, including machine learning, graph theory, and spectral theory and for certain evolution operators, information theory, optimization and also inverse problems methods both variational and statistical inverse problem approaches.